\documentclass[]{interact}

\usepackage[T1]{fontenc}
\usepackage[utf8]{inputenc}
\usepackage{graphicx}
\usepackage{color}
\usepackage[version=4]{mhchem}
\usepackage{verbatim}
\usepackage{subcaption}    %

\usepackage[pdftex,plainpages=false,
    pdfpagelabels,
		pdfusetitle,	
    bookmarksnumbered,
    linkbordercolor={1 1 1},
    hidelinks=true,
    colorlinks={ true },
]{hyperref}

\usepackage[square, numbers]{natbib}
\usepackage{amsmath}
\usepackage{lineno}
\usepackage{geometry}
\usepackage{iftex}
\usepackage{color,soul} %
\usepackage[version=4]{mhchem}

\newcommand{\etal}{et\,al.}
\newcommand{\eg}{e.\,g.}
\newcommand{\ie}{i.\,e.}

\newcommand{\rateU}{\ensuremath{\text{cm}^{3}\,\text{s}^{-1}}}

\newcommand{\COp}{\ce{CO+}}
\newcommand{\HCOp}{\ce{HCO+}}
\newcommand{\COOp}{\ce{CO2+}}
\newcommand{\HCOOp}{\ce{HCO2+}}

\newcommand{\NNp}{\ce{N2+}}

\begin{document}

\articletype{ORIGINAL ARTICLE}%

\title{Cold CAS Ion Trap -- 22 pole trap with ring electrodes for astrochemistry}

\author{
\name{Pavol Jusko\textsuperscript{a$\star$}\thanks{\textsuperscript{$\star$}Contact: pjusko@mpe.mpg.de}, 
Miguel Jim\'{e}nez-Redondo\textsuperscript{a},       
Paola Caselli\textsuperscript{a},
}
\affil{
\textsuperscript{a}{Max Planck Institute for Extraterrestrial Physics, Gießenbachstraße 1, 85748 Garching, Germany}
}
}

\maketitle

\begin{abstract}

The enhancement of a cryogenic radio frequency 22 pole trap instrument by the
addition of ring electrodes is presented in detail. The ring electrodes tightly surround
the poles and only a fraction of the applied electric potential penetrates to the trap
axis, facilitating the fine control of slow cold ions. A precise computational model,
describing the effective mechanical potential created by the applied static and rf
fields, governing the ion behaviour, is employed to demonstrate and understand the
operation of our setup. The use of ring electrodes for improved extraction
of cold stored ions is shown. Variable trapping potentials, placed on one ring
electrode, can be used to control the evaporation of only those \ce{H+} ions from the
trap, whose kinetic energy exceeds the barrier. This ring electrode trapping opens
new possibilities to study processes of minimal kinetic energy release, e. g. spin
exchange. We propose a robust modified method
for the determination of temperature dependent ion molecule reaction rates, resistant
to effects caused by neutral gas freezing and demonstrate it on the reaction of 
\ce{CO+}/\ce{CO2+} with \ce{H2}/\ce{D2}. 
Finally, the use of a supercontinuum laser for quick
localisation of spectroscopic bands is examined on the 
\ce{N2+} 
Meinel system.
\end{abstract}


\section{Introduction}

The 22 pole radio frequency (rf) ion trap was pioneered by Dieter Gerlich more than 30 years ago \cite{Gerlich1992}
as a successor to lower order multipoles, \ie, quadrupoles and octopoles \cite{Teloy1974}. 
$2n$ circular rods (poles) of diameter $d$, easy to manufacture and position, are placed on
a circle with inscribed radius $r_0$. 
These 3 parameters ($2n$, $r_0$ and $d$) are in a first approximation chosen so that the ring circular surface best approximates
the curvature of the hyperbolic potential. For $d=1\;\text{mm}$ and $r_0=5\;\text{mm}$, the magic number ``22'' is born: $2n=22$. 
The use of non-ideal poles and geometries, greatly facilitating the physical manufacturing of the device, 
necessarily leads to perturbations in the ideal field. Fortunately, these perturbations are
mostly not relevant for particle trapping.
This fact, together with the lack of need for bulky and expensive magnets, lead to the wide adoption of radio frequency traps
with many different geometries, from simple linear multipoles, \eg, quadrupole \cite{Willitsch2012, Tesler2018, Miossec2022}, 
hexapole \cite{Kang2005}, octopole \cite{Terasaki2007, Jusko2013},
16 pole \cite{Asmis2002},
22 pole \cite{Stearns2007, Fujihara2008, Mikosch2008, Gerlich2012, Asvany2014, Gunther2017, Kumar2018, Jusko2019, Campbell2020}, 
ring electrode traps (stacked ring electrodes) \cite{Luca2001, Goebbert2009aip},
quadrupole ion traps (3 dimensional quadrupole, QIT) \cite{Wang2008, Wolk2014, Feraud2014} to geometries
where thin wire is used to approximate the desired electrode shape \cite{Jasik2013, Geistlinger2021}.

The main advantage of the 22 pole trap geometry is provided by the high number of poles, leading to 
an almost field free region and sharp steep barriers on the sides of the trap %
approximating a square box potential. 
The high number of poles also absorbs some small manufacturing imperfections as demonstrated by ions still remaining trapped in a 22 pole trap
exhibiting 10 potential minima instead of an ideally predicted flat minimum along the axis \cite{Otto2009} (for symmetry breaking in multipoles see \cite{Pedregosa2017}).
In reality, even the electrostatic potentials at the ends of the axially symmetric trap (used to reflect the ions in 
the axial direction) do affect the potential in the middle of the trap \cite{Willitsch2012,Fanghanel2017},
creating an off-axis potential minimum. %
The effect of these divergences from the ideal potentials become more and more pronounced as the ion temperature (its kinetic energy)
is decreased to cryogenic temperatures (ca. $10\;\text{K}$ or $1\;\text{meV}$ and lower) using an active form of cooling.
In this work, we exclusively focus on buffer gas cooling, \ie, the use of collisions of ions and a cold neutral gas (usually \ce{He}). 
The neutrals are thermalised to the temperature of the inside wall of the trap, as in molecular regime,
the mean free path is greater than the dimensions of the trap, and are not affected by the rf fields. 
The ions, on the contrary, can gain kinetic energy from the rf, which can lead to situations where further decrease of 
the trap body temperature, \ie, of the neutral gas temperature, does not lead to a decrease of the ion temperature, an effect often 
described as parasitic heating ($T_{\text{ion}}>T_{\text{gas}}$). 

The temperature of the ions in the trap can be determined experimentally using chemical probing \cite{Plasil2012}, 
using doppler broadening \cite{Asvany2008, Jusko2014, Brunken2017},
from the rovibrational band profile \cite{Ishiuchi2017, Jusko2019}, 
from hot bands in the electronic spectra \cite{Choi2012},
using Time-of-Flight (ToF) \cite{Notzold2020},
or evaporative ion losses \cite{Lakhmanskaya2014}.
Numerical simulations of kinetic ion temperature were used to understand the discrepancy
between the ion temperature and the trap temperature in a 22 pole trap \cite{Asvany2009} and to
compare the heating effect in a 16 pole and 16 wire pole \cite{Rajeevan2021}. 
The heating effects may also be taken advantage of, \eg, a QIT where a low externally 
induced resonant frequency excitation is used to eject specific $m/q$ ions based on their secular motion \cite{Kang2014}.

The potential inside a multipole can also be influenced by surrounding the multipole with a ring electrode.
Due to the shielding provided by the multipole rods themselves, only a fraction of the applied voltage penetrates
to the multipole axis \cite{Gerlich1992, Fanghanel2017}. Ring electrodes have been used in guided ion beam (GIB)
experiments \cite{Mark1996, Savic2020}, as well as in an octopole trap \cite{Jusko2013}.
Ring electrodes on a 22 pole trap were studied in Dieter Gerlich's group using a rough numerical model (finite difference methods)
and the barrier height was calibrated using partial reflection of ions  stored 
in a packet (pulse) on the ring electrode \cite{Haufler1996}.
Furthermore, Richthofen \etal~\cite{Richthofen2002} used ring electrodes to trap the ions and resonantly excite specific $m/q$ species.
The addition of ring electrodes provides finer access to the trap volume, \eg, to the trap center, which is only slightly affected by the
input/output electrodes, opens the possibility to form barriers inside the trap, \ie, two (or more) separate traps
adjacent to each other, as well as an option to compensate small patch electric fields. Further, the fine tuning of
the potential can be used to influence the ions depending on the energy, \eg, leak out of ions with excess kinetic energy.
The degree of control achieved with this configuration, where ring electrodes act only by penetration, is hard to replicate 
with a ring electrode trap ({\eg} {\cite{Luca2001}}). In the latter case, a very precise determination of the electric potential at the 
electrode surface would be required due to the direct exposure of the ring electrodes to the trap volume.

Ion traps, as demonstrated by all the aforementioned setups, are an extraordinarily versatile tool. 
Their application ranges from fundamental physics through physical chemistry up to molecular biology. 
Of particular interest for us is their application to
laboratory astrochemistry.

Interstellar clouds are cold ($10-100\;\text{K}$) and tenuous ($10^2-10^5$ \ce{H2} molecules 
per $\text{cm}^{3}$) and, in less dense regions, are being impinged by UV photons from the 
interstellar radiation field \cite{Stahler2004}.
Despite these harsh conditions, interstellar clouds are filled with molecules \cite{McGuire2022} 
thanks to the active chemistry initiated by ion-molecule reactions \cite{Herbst1973,Watson72b}. 
Measuring the rates of ion-molecule reactions is crucial for astrochemical models 
(\eg\ \cite{Wakelam15,Sipila15b}), which are used to predict and interpret observations 
of interstellar molecules, important diagnostic tools for the chemical and physical composition of 
interstellar clouds, where stars like our Sun and planets like our Earth form. 
The low particle number densities and cold environments to simulate make ion traps an 
excellent tool for laboratory studies of these astrochemical processes.

In this paper we present a new experimental cryogenic 22 pole rf trap setup Cold CAS Ion Trap (CCIT) 
at The Center for Astrochemical Studies (CAS) at the Max Planck Institute for Extraterrestrial Physics,
which is specifically designed to take advantage of the coupling of ring electrodes to a multipole trap, and its 
application to the study of astrochemical processes such as those occurring in interstellar clouds.

\section{Experimental}

\subsection{Experimental setup}

\begin{figure}[]
\centering
\includegraphics[]{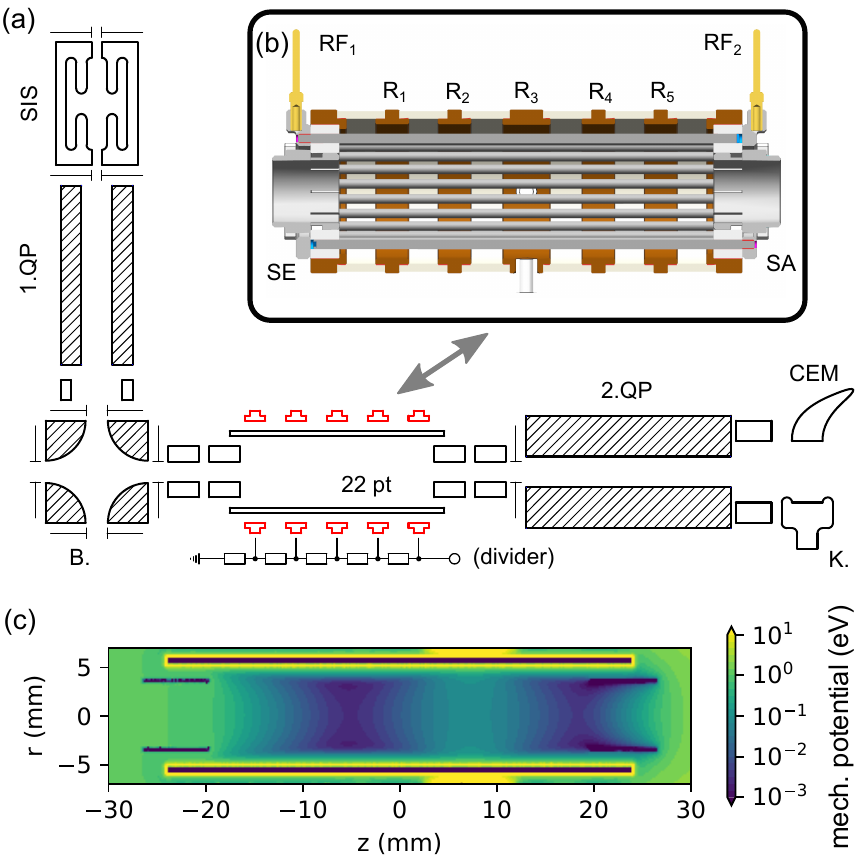}
\caption{Panel (a): Schematic view of the setup. SIS -- Storage Ion Source, 1.QP -- source quadrupole, 
         B. -- electrostatic bender, 22 pt -- 22 pole trap, 2. QP -- product quadrupole, K. - Daly Knob (conversion dynode),
         CEM -- channel electron multiplier.
         Panel (b): 3D model of the 22 pole trap. Panel (c): effective mechanical potential inside the trap
         for a singly charged particle of mass $4\;m/q$, 
         $V_0=50\;\text{V}$, $f_0=19\;\text{MHz}$, $\text{SE}=1\;\text{V}$, 
         $\text{SA}=0\;\text{V}$, $\text{R}_4=100\;\text{V}$, the rest of the ring electrodes are at 0~V. 
  } 
\label{fig:setup}
\end{figure}

The block diagram of the experimental setup can be seen in Fig.~\ref{fig:setup} panel (a). The ions follow a path from the top 
left to bottom right of the diagram accomplished exclusively through the use of electric fields. 
All the measurements proceed in a cyclic manner with periods in multiples of $1\;\text{s}$. Synchronisation is provided by 
computer pre-generated signals with a resolution of $1\;\rm{\mu}\text{s}$.   

The Gerlich type Storage Ion Source (SIS) \cite{Gerlich1992} is used to produce ions. In this type of source,
electron bombardment of a precursor gas continuously leaked in the source produces ions inside a cavity of stacked 
electrodes with an applied rf field facilitating the ion storage. 
The ions are accumulated in the source and are typically extracted only for tens of ms per cycle, thus the storage
function effectively increases the ion yield. Moreover, the produced ions can undergo reactive collisions with neutrals
present in the precursor gas, allowing us to form ions not directly created by electron bombardment, as well as non reactive
collisions leading to the internal relaxation of molecular ions.
We typically produce a wide variety of ions in the SIS and use a source quadrupole (1. QP) to filter only the 
mass of the ion of interest. After the 1. QP the ions are refocused and bent to the trap axis in the electrostatic bender (B.).

The 22 pole trap (22 pt) (see Fig.~\ref{fig:setup} panel (b)) is made out of $1\;\text{mm}$ stainless steel rods on an 
inscribed radius of $5\;\text{mm}$. The rods are held in sapphire rings and every other is electrically connected to 
either of the outputs of the rf generator ($\text{RF}_1$ or $\text{RF}_2$) respectively at both ends.
Input (SE) and output (SA) electrodes are inserted into the trap, ring electrodes $\text{R}_1$--$\text{R}_5$ 
surround the trap rods. The trap is mounted on top of a RDK-101E cryocooler (Sumitomo), and an additional heating element 
HTR-50 (LakeShore) allows us to regulate the temperature down to $4\;\text{K}$. 
Silicon diodes DT-670C-CU (LakeShore) are used to measure the temperature. 
The neutral gas can be administered to the trap 
through two independent lines, allowing either continuous injection or a short pulse using a custom piezo-element actuated 
valve located inside the vacuum chamber close to the thermal shield of the trap unit.
The pressure in the 22 pole trap vacuum vessel is measured by a Bayard-Alpert ionization gauge AIG17G (Arun Microelectronics) which is 
calibrated by a capacitance manometer CMR 375 (Pfeiffer), directly connected to the trap volume and held at room temperature.
Thermal transpiration is taken into account.

The ions leaving the trap are mass selected in the product quadrupole (2. QP), comprising a 
TSQ 7000 hyperquad (Finnigan MAT) driven by QMH 400-5 (Pfeiffer) electronics customised for the particular 
capacitive load of the QP rod system, and subsequently detected in the corresponding TSQ 7000 type detector. 
The system consists of a conversion dynode (knob -- K.) held at 15 kV producing secondary particles (cations, anions, electrons, 
neutrals) upon incoming ion impact. 
In the positive ion mode, the dynode is held at $-15\;\text{kV}$ and the secondary particles are anions and electrons.
When in negative ion mode, the dynode is held at $+15\;\text{kV}$ and the secondary particles are positive ions. Secondary particles 
are detected in a channel electron multiplier (CEM, Photonis 5903 or DeTech 2312), rather than converted to photons as in a typical 
Daly type detector \cite{Daly1960}.
The CEM is used in counting mode and the output signal is discriminated using a model 6908 (Phillips Scientific) discriminator, 
or directly in the multi-channel scaler (MCS) PMS-400A (Becker \& Hickl).

A standard experiment sequence consists of ion production/ion trap filling, storage with exposition 
(ions exposed to photons, neutral reactant, etc.), and product analysis/detection. The cryostat operates with a $1\;\text{s}$ period and 
the experiment is in phase with the cryostat. Additionally, ring electrodes, 22 pole rf amplitude, and MCS 
can be controlled during the cycle in order to acquire particular data.

\subsection{Potentials in a trap with rings}
\label{sec:potentials}

Numerical simulations of the potential generated inside the trap are a useful tool for experimental optimization. 
They have been employed for this purpose by several groups (see for instance  \cite{Fanghanel2017,wilcoxImprovedIonExtraction2002,gibsonModellingMassAnalyzer2010,Rajeevan2021,Otto2009}), either 
through the use of commercial software like SIMION \cite{SIMION}, or open source libraries. Here, we use the boundary 
element method (BEM), implemented through the Python library bempp-cl \cite{smigajSolvingBoundaryIntegral2015,betckeBemppclFastPython2021}, 
to calculate the electrostatic potentials generated by the different trap electrodes. 
BEM calculations avoid the need for a spatial grid in which the solution is calculated, 
allowing instead the evaluation of the potential at any point in space, 
and the use of a Python library facilitates 
the analysis of the computed results and the application of the calculated potential in
further simulations. The solution of the Laplace equation is computed 
using the fast multipole method \cite{greengardFastAlgorithmParticle1987}, which simplifies long distance interactions between 
discrete elements lowering the memory requirements for the simulation. The boundary conditions are imposed on a 3D mesh 
generated from a 3D CAD model of the 22 pole trap using the Salome platform \cite{SalomePlatform}.

Given a general electric potential inside the trap with rf and static components
\begin{equation}
    \Phi = \Phi_{\mathit{rf}} \cos \left( \Omega t \right) + \Phi_{\mathit{s}} \; ,
\end{equation}

where  $\Omega = 2 \pi f_0$ is the rf angular frequency, the average force acting on an ion with mass $m$ and charge $q$ can be derived from the effective 
mechanical potential \cite{Gerlich1992}
\begin{equation}
        V^{*}= \frac{q^2 E_{\mathit{rf}}^2}{4 m \Omega^2} + q\Phi_{\mathit{s}} \; ,
\end{equation}
where $\mathbf{E}_{\mathit{rf}}$ is the amplitude of the rf electric field, obtained as
\begin{equation}
    \mathbf{E}_{\mathit{rf}} = - \nabla \Phi_{\mathit{rf}} \; .
\end{equation}

This approach requires a separation of the ion motion into a slow drift term, controlled by the effective potential, and 
a rapid oscillatory motion due to the rf field. To test the validity of this approximation, the adiabaticity parameter 
$\eta$ can be defined as
\begin{equation}
\eta = \frac{2 q \left| \nabla E_{\mathit{rf}} \right| }{m \Omega^2} \; .
\end{equation}

A value of $ \eta < 0.3 $ generally guarantees that the approximation holds.

An analytical solution exists for the case of an ideal multipole of order $n$. The effective potential and adiabaticity 
parameter in plane polar coordinates $\left( r,\varphi \right)$ then take the form
\begin{equation}
    V^* = \frac{n^2 q^2 V_0^2}{4 m \Omega^2 r_0^2} \left( \frac{r}{r_0} \right)^{2n-2} + 
    q U_0 \left( \frac{r}{r_0} \right)^{n} \cos \left( n \varphi \right)
    \label{eq:veff}
\end{equation}
\begin{equation}
    \eta = 2 n \left( n-1 \right) \frac{q V_0}{m \Omega^2 r_0^2} \left( \frac{r}{r_0} \right)^{n-2} \; ,
    \label{eq:eta}
\end{equation}
where $\Phi_0 = U_0 - V_0 \cos \left( \Omega t \right)$ is the potential applied to the electrodes, and $r_0$ 
is the inscribed radius.

The computational method described above has been used to obtain the effective potential inside the trap for 
different experimental configurations, such as the one depicted in panel (c) of Fig.~\ref{fig:setup}. In this 
particular case, the rf field is generated with the 22 poles and the axial trapping of the ions is performed 
by the end electrode at the entrance of the trap, SE, and the fourth ring electrode $\text{R}_4$. Since ring 
electrodes only change the trap potential by penetration, the 100 V voltage applied to the electrode only 
results in a barrier of tens of meV inside the trap (see below).

\begin{figure}[h]
\centering
\includegraphics[]{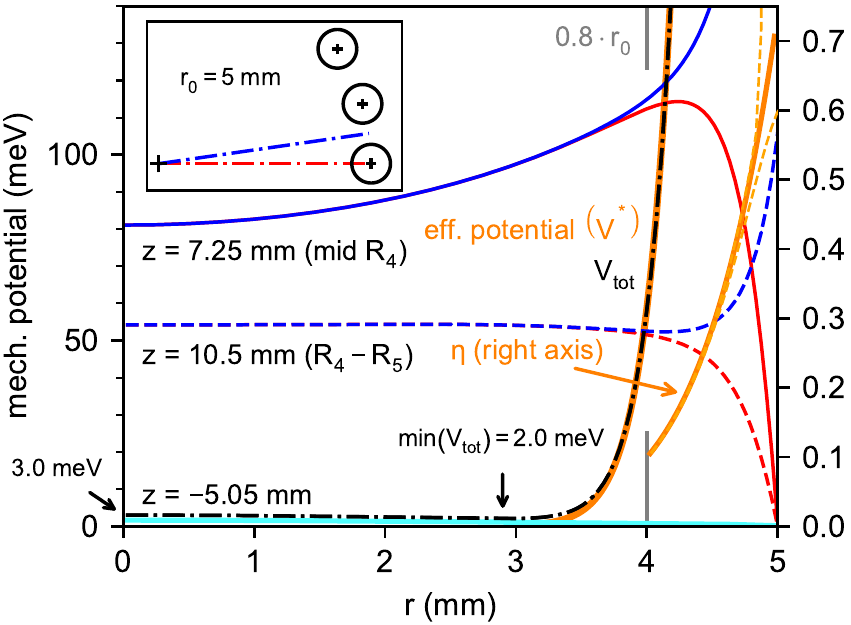}
\caption{Radial profile of different potentials in the trap. In red and blue, potential generated by the $\text{R}_4$ 
ring electrode at different $z$ positions along the axis of the trap, towards the center of the rod (red) and towards 
the middle of the gap between rods (blue); cf. top left inset. 
Cyan line for $z=-5.05\;\text{mm}$ represents the overlap of blue and red lines. 
Black dashed-dotted line: effective potential at $z=-5.05\;\text{mm}$ for the trap configuration described in Fig. \ref{fig:setup}. 
Arrow marks the minimum of the total potential at $r= 2.9\;\text{mm}$. Solid orange lines: analytical expressions for the effective 
potential and adiabaticity parameter for the corresponding ideal multipole. Dashed orange lines: calculated adiabaticity 
parameter towards the center of the rod and the middle of the gap between the rods (shown only for $r>0.8\cdot r_0$).  } 
\label{fig:radial}
\end{figure}

A detailed look at the radial profile of the potential inside the trap is included in Fig. \ref{fig:radial}. 
The potential generated by just the $\text{R}_4$ electrode at different positions, as well as the effective 
potential and adiabaticity parameters, are shown. The analytical values from eqs. \ref{eq:veff} and \ref{eq:eta} 
are also plotted for comparison. It can be seen that the radial profile of the $\text{R}_4$ potential changes 
depending on the position $z$ along the axis of the trap. At the position of the ring electrode ($z=7.25$ mm), 
the potential barrier has its minimum at the axis, and gradually increases with the radius. At $z=10.5$ mm 
(between the fourth and fifth ring electrodes), the shape changes and now the potential is almost constant from 
the axis up to $r \sim 3$ mm, with a slight drop between 3 and 4 mm. In both cases, differences in the potential 
towards a rod or the gap between them are only significant for $r > 4$ mm, with both potentials quickly diverging 
towards 0 and $\sim 100$ eV respectively. The potential generated by $\text{R}_4$ quickly decays away from 
the ring electrode, and at $z=-5.05$ mm, where the minimum potential from Fig. \ref{fig:setup} (c) is found, its value 
is $\sim 1$ meV at the axis, decreasing towards the poles. Because of this radial shape, which is similar to the one 
generated by the end electrodes \cite{Fanghanel2017}, the calculated effective potential for the complete configuration 
has its minimum $\sim 3$ mm off the axis and towards the rods. The calculated effective potential matches the analytical 
expression from eq. \ref{eq:veff} closer to the rods, where the effect of the multipole field surpasses that of the end 
and ring electrodes. A similar result is obtained for the adiabaticity parameter from eq. \ref{eq:eta}.

\begin{figure}[h]
  \centering
  \includegraphics[]{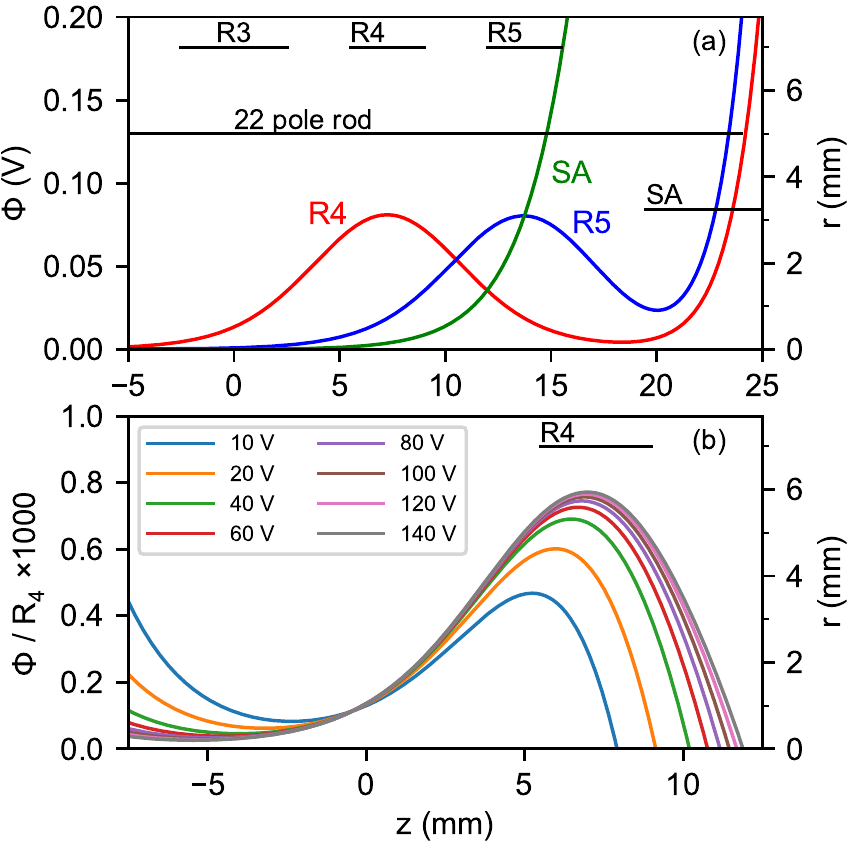}
  \caption{Panel (a): Axial profile of the electrostatic potentials generated by the two ring electrodes closest to the 
    trap exit, $\text{R}_4$ and $\text{R}_5$, when set to 100 V, and the output electrode SA set to 1 V (see Fig.~\ref{fig:setup}).
    Panel (b): Axial profile of the relative electrostatic potentials generated by the $\text{R}_4$ ring electrode with 
    respect to the applied voltage, when the output electrode is set to an extraction voltage of $\text{SA}=-1\;\text{V}$.
  }\label{fig:relative}
\end{figure}

The combination of potentials generated by ring and end electrodes should be carefully undertaken, since they can interfere 
with one another as noted by Fanghänel et al. \cite{Fanghanel2017}. Panel (a) of Fig. \ref{fig:relative} shows the axial 
profile of the electrostatic potentials generated by the $\text{R}_4$ and $\text{R}_5$ ring electrodes and the output electrode SA. 
As can be noted, the potentials of $\text{R}_5$ and SA clearly overlap with each other, so combinations of $\text{R}_4$ and 
SA should be preferred. A useful configuration for ion energy sampling (see Discussion \ref{sec:disc}) is to set a potential 
barrier with $\text{R}_4$ and extract the ions that go over it using a negative potential in SA. The axial profile of 
the combined potential generated for that configuration is shown in Fig. \ref{fig:relative} (b), normalized to the potential 
applied to the $\text{R}_4$ electrode. The curves for the different voltages do not overlap, which means that the influence 
of SA distorts the potential barrier somewhat so that its height is no longer proportional to the applied voltage. 
Above 40 V, the curves start to get closer as the contribution of SA becomes relatively smaller, and eventually converge to a corresponding penetration of $\sim 8 \cdot 10^{-4}$.

\subsection{Applications of ring electrodes}

\begin{figure}[h]
  \centering
  \includegraphics[]{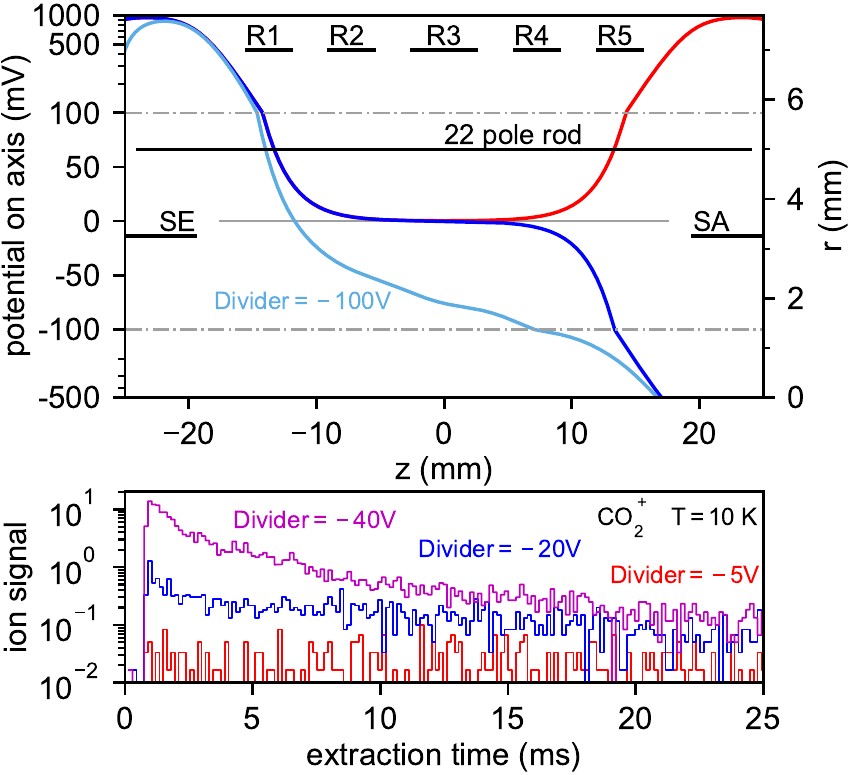}
  \caption{Trapping and emptying of the trap equipped with ring electrodes.
    Top panel: axial profile of electrostatic potential produced by the electrodes.
    Red colour represents trapping potential, dark blue colour represents emptying 
    potential with no ring electrodes. The use of ring electrodes and the divider (see Fig.~\ref{fig:setup})
    creates a uniform decreasing potential (light blue). 
    Bottom panel: Effect of the divider on the extraction of mass $44\;m/q$ ions from the trap held at $10\;\text{K}$.
  }\label{fig:axial}
\end{figure}

The ring electrodes offer fine tuning of the ion conditions inside the ion trap. Applying the right potential, the 
position of the ion cloud can be influenced or ions can be directly trapped using only ring electrodes \cite{Richthofen2002}. 
The ion cloud can also be pushed out of the trap while emptying the trap as shown in Fig.~\ref{fig:axial}. 
In some experimental conditions (trap temperature $<10\;\text{K}$, high neutral number densities) it may 
become difficult to empty the trap only by using the end electrodes SE, SA. 
We believe this is mainly caused by the fact that the induced emptying axial potential (dark blue in 
Fig.~\ref{fig:axial}) is monotonously decreasing, albeit very slowly, due to shielding of the SA electrode 
by the trap rods themselves. 
The emptying potential can be made an order of magnitude steeper just by using a tens of volts lower potential
(in the case of cations) on every consecutive ring electrode (see divider in Fig.~\ref{fig:setup}), 
effectively overcoming possible patch potentials on the trap rods as well as potentials present due to manufacturing
imperfections.

\section{Results} %

\subsection{\ce{H+} evaporation from the trap}
\label{sec:evap}

\begin{figure}[h]
\centering
  \includegraphics[]{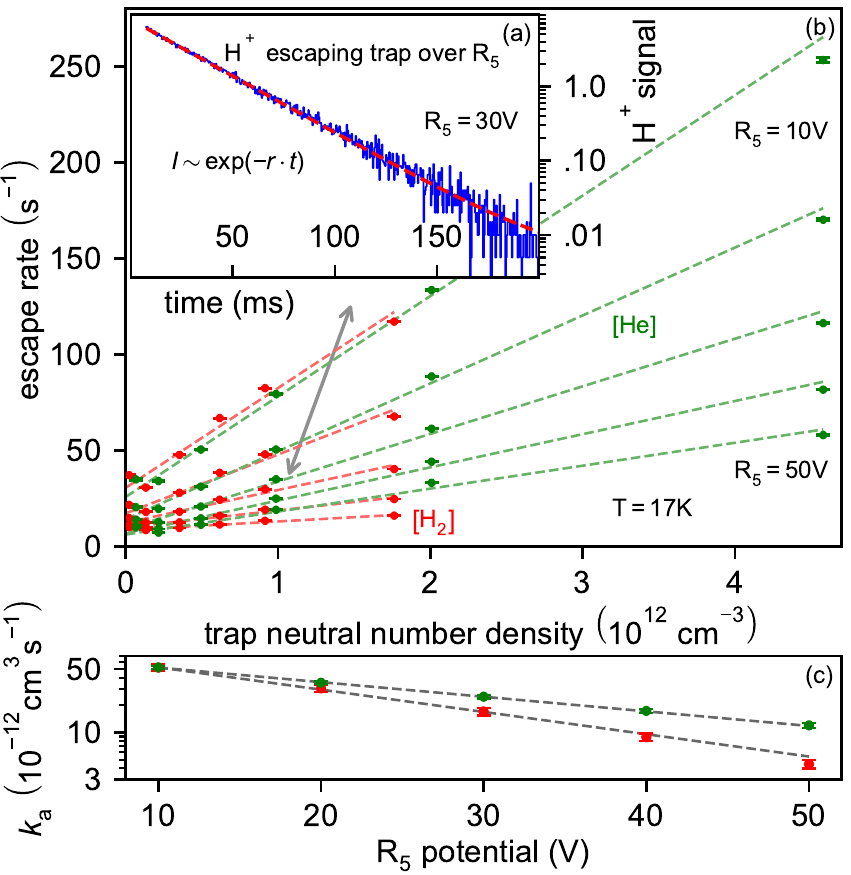}
  \caption{Evaporation of \ce{H+} from the trap over $\text{R}_5$ electrode as a function of electrode potential, 
    number density and neutral gas. Inset (a): evaporation (escape rate $r$) from the 22 pole trap as seen with the MCS.
    Panel (b): escape rate $r$ as a function of neutral number density for \ce{H2}, \ce{He} gas 
    (gray arrow marks the point corresponding to inset (a)).
    Panel (c): Apparent collisional rate $k_{\text{a}}$ as a function of the potential on the $\text{R}_5$ electrode.
  } 
\label{fig:evap}
\end{figure}

The low potential barrier created by a ring electrode can be used to examine processes in which the energy of the trapped 
ions plays a significant role. In this experiment we intentionally keep the barrier formed by ring electrode $\text{R}_5$ as low as possible
in order to quantify the escape of \ce{H+} ions in different conditions.
\ce{H+} ions are first trapped and cooled down using initial \ce{He} pulse and SE, SA electrodes. Subsequently,
after the He buffer gas is pumped away, the SA barrier is removed and only the $\text{R}_5$ electrode
is able to reflect the ions back to the trap. Ions that are not reflected pass over the barrier and are detected
(see Fig.~\ref{fig:evap}(a)). This ``evaporation'' process is of exponential nature over several orders of magnitude
and can be represented by a single number, the escape rate $r$. We plot the escape rate $r$ at various number densities,
neutral gases (\ce{H2}, \ce{He}) and $\text{R}_5$ electrode potential in Fig.~\ref{fig:evap}(b).
It is immediately clear, that $r$ is linear with number density, implying a binary process and can be characterised
by an apparent collision rate $k_{\text{a}}$ (Fig.~\ref{fig:evap}(c)).

Comparing the $k_{\text{a}}=5.2\cdot10^{-11}\,\rateU$ (at $\text{R}_5=10\;\text{V}$) to the Langevin collisional rate 
$k_{\text{L}} = 2.6\cdot10^{-9}\,\rateU$ for $\ce{H+} + \ce{H2}$ reveals that approximately only $2\,\%$ 
of collisions lead to evaporation (for $\ce{H+} + \ce{He}$ $k_{\text{L}} = 1.2\cdot10^{-9}\,\rateU$ 
implying cca. $4\,\%$ evaporation). 
The $k_{\text{a}}$ decreases faster for \ce{H2} than for \ce{He} with increasing $\text{R}_5$ voltage 
(Fig.~\ref{fig:evap}(c)), \ie, collisions with \ce{H2} are clearly different than collisions with \ce{He}. 
On top of the apparent difference in the target mass (factor of 2), one has to also consider that the
collision with \ce{He} is completely non-reactive as is the case for a collision with para-\ce{H2}.
Since normal room temperature \ce{H2} is used, the corresponding 3:1 ortho to para ratio in \ce{H2} is assumed.
In the case of a collision with ortho-\ce{H2}, the spin conversion reaction 
\begin{equation}
\ce{H+} + \ce{H2}(o) \to \ce{H+} + \ce{H2}(p) 
\label{eq:h+h2}
\end{equation}
occurs with a predicted reaction rate $k_{\text{o-p}}\approx 1-2\cdot10^{-10}\,\rateU$ 
\cite{Gerlich1990, Honvault2012, Grozdanov2012, Lezana2021} 
around $10\;\text{K}$ and releases $14.7\;\text{meV}$ \cite{Beyer2019} of kinetic energy,
which is mostly carried away in the lighter \ce{H+} ion ($\approx10\;\text{meV}$) due to conservation of momentum. 
This additional energy greatly exceeds the thermal energy ($<2\;\text{meV}$ at $17\;\text{K}$) 
available in non-reactive collisions and contradicts our initial expectation, where \ce{H2} would have been
responsible for a stronger escape effect.

To our knowledge, this is the first attempt to characterise 
the reaction~\ref{eq:h+h2}, where the reactants and products do not change their $m/q$ ratios
and the reaction leads only to a minimal energy release.
Our initial assumption, that the \ce{H+} ion will be less affected by the collision with neutral \ce{He}
atoms, than with the \ce{H2} molecules, where the reaction is possible was found to be false (see Fig.~\ref{fig:evap}(b)).
We presume, that within the current experimental framework, the reaction can only be studied  by 
using highly enriched para-\ce{H2} as the reference measurement and attribute the 
reaction~\ref{eq:h+h2} to the difference to the 
measurement with normal-\ce{H2} (contains 3:1 ortho:para \ce{H2}) instead of directly using normal-\ce{H2}.

\subsection{Reaction rates}
\label{sec:k}

All ion traps, irrespective of being based on rf or magnetic fields, offer considerable trapping times making
them prefered tools for ion-neutral interaction studies, from fast near Langevin processes, typically $10^{-9}\;\rateU$, to 
slow radiative attachment, where the reaction rate coefficient is in the order of $10^{-16}\;\rateU$ \cite{Gerlich2013}.

\begin{figure}[h]
\centering
  \includegraphics[]{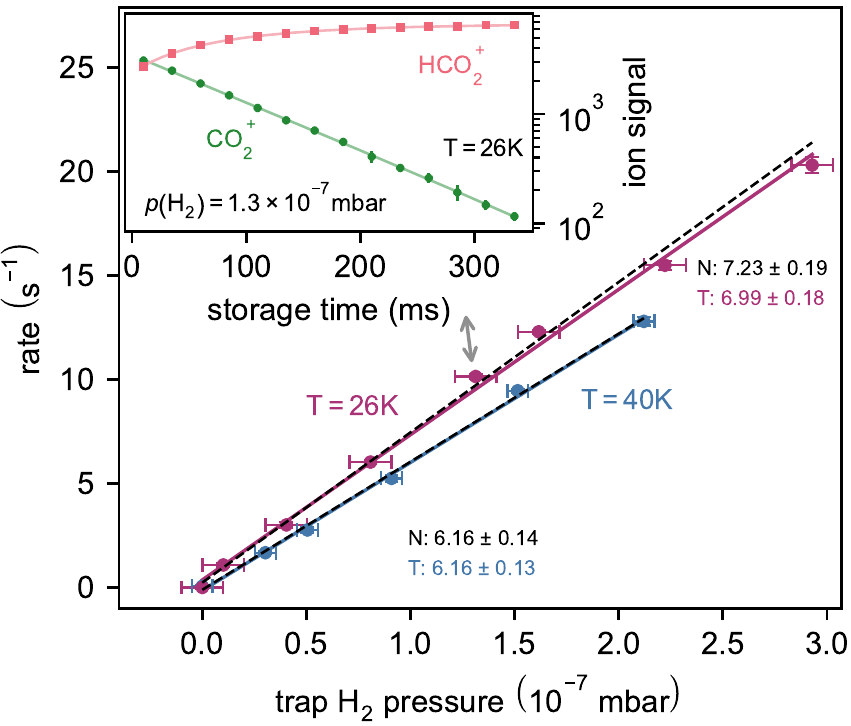}
  \caption{
  Reaction rate of $\COOp + \ce{H2} \to \HCOOp + \ce{H}$ as a function of \ce{H2} pressure at trap temperatures of 26 and $40\;\text{K}$. 
  Inset: Number of ions in the trap as a function of storage time.
  Loss rate in absence of reactants is negligible at time scales of 
  hundreds of ms.
  Grey arrow points to the data point, which is represented by the inset. N, T -- normal LS, total LS respectively (see text).
  } 
\label{fig:slope}
\end{figure}

The reaction rate coefficient determination procedure, illustrated on the reaction
\begin{equation}
\COOp + \ce{H2} \to \HCOOp + \ce{H}
\label{eq:coo}
\end{equation}
can be seen on the inset of Fig.~\ref{fig:slope}. 
The ion trap is filled with the reactant ion \COOp, while a known pressure of neutral \ce{H2} is maintained inside the trap.
The product of the reaction, \HCOOp, accumulates in the trap as the number of reactant \COOp\ ions decreases 
(neutral \ce{H} product can not be stored/detected). 

This binary process can be described as exponential loss of the primary ion \COOp. 
Every point in the inset of Fig.~\ref{fig:slope} was recorded 10 times (see the error bars), the ``ion signal'' (abscisa)
is always reported in numbers per filling throughout the whole document. The reaction rate coefficient at this given temperature can
simply be determined as the division of the least square (LS) fitted time constant and neutral number density.
Usually, this procedure is repeated over decreasing temperature, as the cryostat is being cooled down. 
The whole process is usually repeated in order to bin or average the measured reaction coefficients and arrive at a more
robust final result.

We want to report a different approach, usually only used to test the order of reaction, where the reaction rate
is measured in a wide range of number densities of the neutral reactant at every temperature (Fig.~\ref{fig:slope}).
The slope of this dependence is the reaction rate coefficient $k$. The solution of this overdetermined system is best found using
regression analysis in the form of a least squares fit. Contrary to the inset case, where the ordinate value (time) has 
negligible uncertainty, the pressure (or number density, etc.) can have considerable scatter, therefore, we investigated the use of LS methods
taking both errors into consideration \cite{Tellinghuisen2020}. 
Fig.~\ref{fig:slope} compares ordinary least squares (dashed line; number ``N''; ignores the pressure uncertainty) and 
total least squares (full line; number ``T''). Although the difference in the fitted slope $k$ is within the fit uncertainty even for the 
$T=26\;\text{K}$ measurement data set, where the pressure uncertainty is relatively high, we recommend the use of total LS
for $k$ determination. The method will provide more consistent results, especially in the case where the pressure uncertainty is not 
even over the measurement range.

\begin{figure}[h]
  \centering
  \includegraphics[]{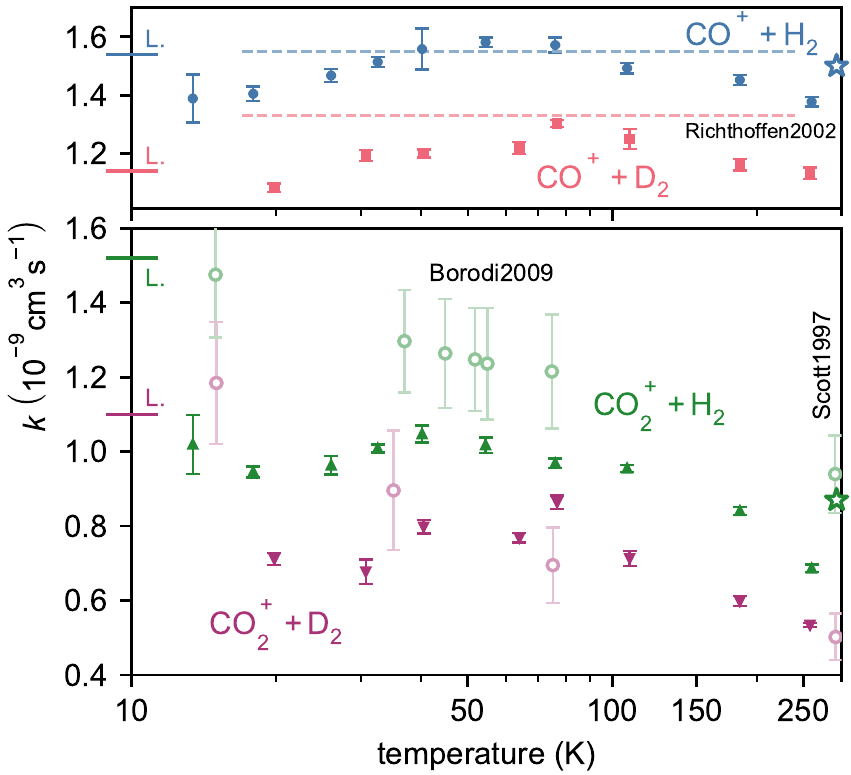}
  \caption{Reaction rate coefficient of \COp\ and \COOp\ with \ce{H2} and \ce{D2} as a function of temperature.
    Langevin rate coefficient shown on the left (dash; L.). 
    Previous results are reported using open symbols and dashed lines 
    (\COp\ \cite{Scott1997, Freeman1987, Richthofen2002} and \COOp\ \cite{Borodi2009, Scott1997}).
  } 
\label{fig:ks}
\end{figure}

The temperature dependent reaction rate coefficient of the reaction 
\begin{equation}
\COp + \ce{H2} \to \HCOp + \ce{H}
\label{eq:co}
\end{equation}
and reaction \ref{eq:coo} is shown in Fig.~\ref{fig:ks}.

We confirm the increase of the reaction rate with decreasing temperature for the reaction of $\COOp+\ce{H2}$ (and \ce{D2}),
while providing more data points with superior statistical error that rather show a leveling-off at cryogenic temperatures
below $40\;\text{K}$, in contrast with the results of Borodi {\etal}~{\cite{Borodi2009}}.
We confirm that the reaction of $\COp+\ce{H2}$ (and \ce{D2}) is mostly constant from $15$ to $250\;\text{K}$ with essentially
Langevin reaction rate coefficient (contrary to Richthofen \etal~\cite{Richthofen2002}, who report the reaction 
rate coefficient with \ce{D2} $15\,\%$ higher, though still within their reported error margin).

The reported error bars always correspond to the standard deviation of the data fit.  
The main source of overall error is the density calibration uncertainty, which we estimate to be in the range $20-30\,\%$. 
At the same time, the error of the relative difference between the measured rate coefficients is not affected by 
the density calibration uncertainty, and shall be close to the statistical one.
We assume the ion temperature is very close to the temperature of the neutrals for $T>15\;\text{K}$ and for the trap parameters employed, 
with low rf amplitudes and $\eta \ll 0.1$ (eq.~\ref{eq:eta}) \cite{Jusko2014, Brunken2017}.

\subsection{Action spectroscopy}

As in any other ion trapping experiment, the number density of ions in our setup is 
considerably lower than $10^5\;\text{cm}^{-3}$, enabling only the use of action type of spectroscopy methods, 
where the ``change'' of the studied medium, rather than that of the absorbed/ emitted photons is observed. 
Action schemes from laser induced charge transfer LICT \cite{Schlemmer1999} 
to pre-dissociation of weakly bound clusters, pioneered in the 1980s \cite{Okumura1985}, are possible.

We report the use of bright supercontinuum laser for overview action spectroscopy of \ce{N2+},
where charge transfer reaction 
\begin{equation}
\ce{N2+} + \ce{Ar} \to \ce{Ar+} + \ce{N2} ~~~\Delta H = 0.18\;\text{eV}
\end{equation}
proceeds predominantly for excited \ce{N2+} states, because of the $0.18\;\text{eV}$ endothermicity.

\begin{figure}[h]
  \centering
  \includegraphics[]{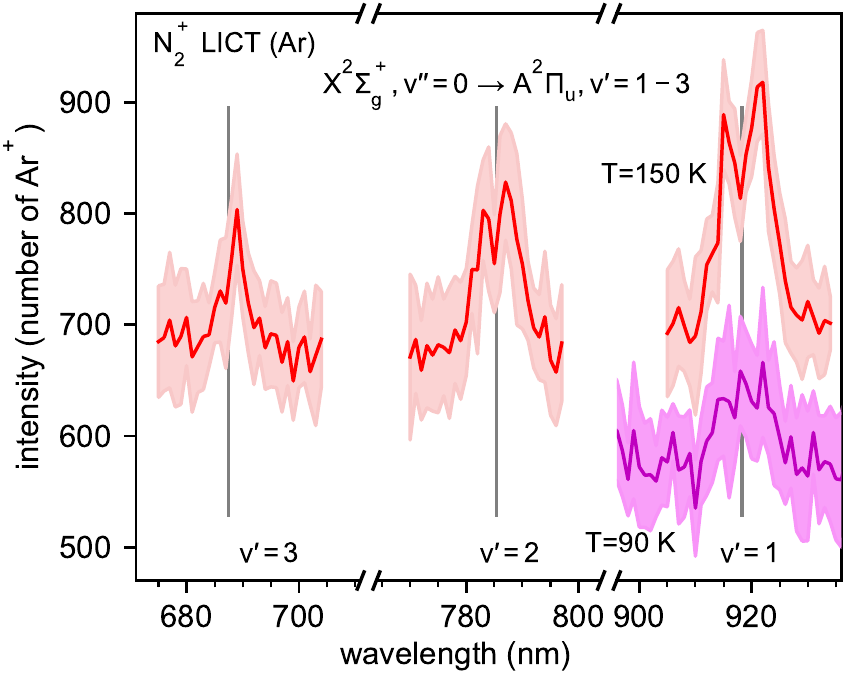}
  \caption{Electronic spectrum of \NNp\ cation Meinel system \cite{Benesch1980} (gray line -- band origins) 
    recorded using VIS-LICT (laser induced charge transfer) scheme \cite{Schlemmer1999}. In red, measurements performed at 150 K. In magenta, at 90 K.
    }\label{fig:lir}
\end{figure}

Light source is SuperK FIANIUM FIU-15 with high resolution bandpass filter LLTF CONTRAST (NKT Photonics)
with spectral range 400 -- 1000 nm (channel spectral bandwidth $< 2.5$ nm FWHM).
A similar laser system has been used previously in \ce{He} tagging overview action spectroscopy 
of PAHs \cite{Roithova2019}, where many spectral features of complex molecule guarantee a 
dense spectrum. This is not the case for a diatomic \ce{N2+} molecule with only the 
lowest rotational states populated in our temperature range, implying a sparse spectrum.   
Fig.~\ref{fig:lir} shows the recorded spectra with better S/N ratio at higher trap temperature ($T=150\;\text{K}$) and signal diminishing 
below $T=90\;\text{K}$. We expect this behaviour to be directly related to the temperature 
dependent \ce{N2+} state population and decreasing kinetic velocity Doppler line broadening.
Unfortunately, no pre-dissociation spectroscopy (\ce{He} tagging \cite{Brummer2003}), nor laser induced inhibition 
of cluster growth (LIICG) \cite{Chakrabarty2013} attempts were successful.

For illustration, spectral power density inside our trap was close to $1\;\text{mW}\,\text{nm}^{-1}$, 
whereas Schlemmer \etal~\cite{Schlemmer1999} achieved around $30\;\text{W}\,\text{nm}^{-1}$. %
The primary advantage of using supercontinuum sources in overview action spectroscopy is the high scanning speed, with the ability to acquire 
the presented spectral features in tens of minutes. 

\section{Discussions}
\label{sec:disc}

We have shown the marginally improved method to investigate the temperature dependence of reaction rate coefficients
in section~\ref{sec:k}. The fundamental source of errors and experimental difficulties remains the control and determination of the
neutral gas pressure (number density) present inside the trap. 
The interaction of the neutral species with the trap walls is particle as well as temperature dependent, most commonly observed as
disappearance of the molecules, \ie, ``freezing''. On one hand, to minimize this effect, we constructed our trap with walls just around the
rods, minimising the trap volume/surface area (see Fig.~\ref{fig:setup}(b)). On the other hand, the trap construction required the use of electrically 
insulating material (ceramics) with unknown adsorption surface properties.
In any case, the application of the method shown in Fig.~\ref{fig:slope} minimises the space for errors, as any discrepancy in the
linearity of the data not only as a function of neutral number density, but inherently also as a function of 
trap temperature and deposition time 
(since measurements take several hours) is immediately seen and its cause can be investigated.

The use of ring electrodes to improve the extraction efficiency from the trap has been demonstrated in Fig.~\ref{fig:axial}
and has been found more and more useful, as the mass of the ion increased (over $20\;m/q$; \ce{H+}, \ce{He+} are virtually unaffected) 
at temperatures $T<50\;\text{K}$ and higher neutral number densities ($10^{12}\;\text{cm}^{-3}$). 
It is intrinsically difficult to investigate the extraction efficiency in our setup, since the measured quantity 
is always a combination of extraction, focusing effects of all the ion optics and second quadrupole, and the dead 
time of the detector. Still, all these elements have to be optimised, 
on case by case basis, for the best extraction efficiency in order to increase the amount of acquired data per time.  
Nevertheless, the improved extraction in the case of the $\ce{CO2+}+\ce{H2}$ experiment, as shown in Fig.~{\ref{fig:axial}} (bottom panel),
sped up data acquisition by more than an order of magnitude and undeniably reinforced the validity of data by decreasing the 
dependence of extraction efficiency on temperature. It is also important to note, that experiments where the trap temperature is
not varied  (\eg, virtually all the spectroscopy experiments), do not suffer from extraction efficiency issues by design,
as it stays constant in the time frame of the signal acquisition.

The evaporation, as described in section~\ref{sec:evap}, depends on the neutral number density but also on the type of the 
neutral particle. Potentially, this fact may be further exploited as a sensitive number density 
(pressure) gauge and used to calibrate the number density in the trap, without the need for a 
calibration reaction, \eg\ the calibration of a neutral molecular beam passing through the trap \cite{Borodi2009}.   

On top of the evaporation, more generally, ring electrodes may also be used to analyse the ions in the trap, 
by configuring the field in a way such 
that only ions fulfilling predetermined conditions (\ie, kinetic energy) will pass the barrier. 
In this way the barrier height may be calibrated using ion beams of defined energy \cite{Richthofen2002}.
Another possible application is recently introduced ``leak-out spectroscopy'' scheme, where only ions affected
by light can leak through the barrier \cite{Schmid2022}.
The potential barrier may also be used to detect the ions undergoing a collision (see section~(\ref{sec:evap}))
or to characterise the energy distribution of ions inside the trap.

\normalsize

\section{Conclusion}

We present a 22 pole trap instrument equipped with five ring electrodes specifically designed to 
study ion-molecule interactions relevant for astrochemistry.
Low temperature interaction with neutrals other than \ce{He} or \ce{H2} (or generally at temperatures below $15\;\text{K}$) remains 
challenging due to cryo-pumping (``freeze out''). We outline a method to determine the temperature 
dependent reaction rate coefficient resilient to errors arising therefrom.
We present a spectroscopic overview approach suitable for fast localisation of spectroscopic features
based on a supercontinuum laser.
We introduce a trap evaporation scheme usable to characterise processes involving spin change,
releasing only minimal kinetic energy, while maintaining $m/q$ composition of trapped particles. 
Further experimental and theoretical work is necessary in order to quantify the reaction
rate of the process from the evaporation rate(s).

We present a detailed computational model used to obtain the effective potentials inside the trap
as a function of voltages applied to all the relevant electrodes and explain how to obtain the 
desired field configuration for injection, trapping, evaporation, and improved extraction from the trap.
The most important conclusion is the fact that the contributions from all the trap electrodes to 
the effective potential have to be considered simultaneously, \ie,
one can not consider only the influence of ring electrode(s) while neglecting the end electrode(s), and that 
the barrier height has to be related to the real (non zero) potential minima inside the trap.
We specifically abstain from any quantitative characterisation of ion energy as a function potential
applied to any electrode and rather focus on relative measurements only (change of \ce{H+} evaporation rate).
The quantitative analysis and associated model, requiring dynamical ion trajectory simulations, will be 
published separately. 

The ability to store ions unobstructed for long times emphasizes the usefulness of cryogenic 
rf ion traps for the characterisation of ions through spectroscopy and studies of ion-molecule interactions 
using reaction kinetics.

\section*{Acknowledgements}

The completion of this work is dedicated to Prof. Dieter Gerlich, 
who started to build the setup with us in 2019 and continued the works until
he passed away in the fall of 2020 (removed from the author list as requested by the editor on 2023-2-8).
The authors gratefully acknowledge the work of the electrical and mechanical workshops and engineering 
departments of the Max Planck Institute for Extraterrestrial Physics.
We thank Prof. Stephan Schlemmer for helpful discussions.

\section*{Disclosure statement}
The authors report there are no competing interests to declare.

\section*{Additional information}
The raw experimental data and corresponding post-processing scripts can be downloaded at \url{https://doi.org/10.5281/zenodo.7410333}.

\section*{Funding}
This work was supported by the Max Planck Society.


\begin{thebibliography}{99}
{}
\bibitem{Gerlich1992}
D. Gerlich, \emph{Adv. Chem. Phys.: State-Selected and State-to-State
Ion-Molecule Reaction Dynamics}, Vol. LXXXII, edited by C.-Y. Ng and
M. Baer (Wiley, New York, 1992), pp. 1–176, doi:\href
{https://doi.org/10.1002/9780470141397.ch1} {\nolinkurl
{10.1002/9780470141397.ch1}}
{}
\bibitem{Teloy1974}
E. Teloy and D. Gerlich, Chemical Physics \textbf{4}, 417–427 (1974).
doi:\href {https://doi.org/10.1016/0301-0104(74)85008-1} {\nolinkurl
{10.1016/0301-0104(74)85008-1}}
{}
\bibitem{Willitsch2012}S. Willitsch, International Reviews in Physical Chemistry \textbf{31}, 175–199
(2012). doi:\href {https://doi.org/10.1080/0144235X.2012.667221} {\nolinkurl
{10.1080/0144235X.2012.667221}}
{}
\bibitem{Tesler2018}
L. F. Tesler, A. P. Cismesia, M. R. Bell, L. S. Bailey, and N. C. Polfer, Journal
of the American Society for Mass Spectrometry \textbf{29}, 2115–2124 (2018).
doi:\href {https://doi.org/10.1007/s13361-018-2026-7} {\nolinkurl
{10.1007/s13361-018-2026-7}}
{}
\bibitem{Miossec2022}
C. Miossec, M. Hejduk, R. Pandey, N. J. A. Coughlan, and B. R. Heazlewood,
Review of Scientific Instruments \textbf{93}, 033201 (2022). doi:\href
{https://doi.org/10.1063/5.0080458} {\nolinkurl {10.1063/5.0080458}}
{}
\bibitem{Kang2005}
H. Kang, C. Jouvet, C. Dedonder-Lardeux, S. Martrenchard, C. Charrière,
G. Grégoire, C. Desfrançois, J. P. Schermann, M. Barat, and J. A. Fayeton,
The Journal of Chemical Physics \textbf{122}, 084307 (2005). doi:\href
{https://doi.org/10.1063/1.1851503} {\nolinkurl {10.1063/1.1851503}}
{}
\bibitem{Terasaki2007}
A. Terasaki, T. Majima, and T. Kondow, The Journal of Chemical Physics
\textbf{127}, 231101 (2007). doi:\href {https://doi.org/10.1063/1.2822022}
{\nolinkurl {10.1063/1.2822022}}
{}
\bibitem{Jusko2013}
P. Jusko, Š. Roučka, R. Plašil, and J. Glos\'{i}k, International Journal of Mass
Spectrometry \textbf{352}, 19–28 (2013). doi:\href
{https://doi.org/10.1016/j.ijms.2013.08.001} {\nolinkurl
{10.1016/j.ijms.2013.08.001}}
{}
\bibitem{Asmis2002}K. R. Asmis, M. Brümmer, C. Kaposta, G. Santambrogio, G. von Helden,
G. Meijer, K. Rademann, and L. Wöste, Phys. Chem. Chem. Phys. \textbf{4},
1101–1104 (2002). doi:\href {https://doi.org/10.1039/B111056J} {\nolinkurl
{10.1039/B111056J}}
{}
\bibitem{Stearns2007}
J. A. Stearns, S. Mercier, C. Seaiby, M. Guidi, O. V. Boyarkin, and
T. R. Rizzo, Journal of the American Chemical Society \textbf{129},
11814–11820 (2007). doi:\href {https://doi.org/10.1021/ja0736010} {\nolinkurl
{10.1021/ja0736010}}
{}
\bibitem{Fujihara2008}
A. Fujihara, H. Matsumoto, Y. Shibata, H. Ishikawa, and K. Fuke, The Journal
of Physical Chemistry A \textbf{112}, 1457–1463 (2008). doi:\href
{https://doi.org/10.1021/jp709614e} {\nolinkurl {10.1021/jp709614e}}
{}
\bibitem{Mikosch2008}
J. Mikosch, R. Otto, S. Trippel, C. Eichhorn, M. Weidemüller, and R. Wester,
The Journal of Physical Chemistry A \textbf{112}, 10448–10452 (2008).
doi:\href {https://doi.org/10.1021/jp804655k} {\nolinkurl
{10.1021/jp804655k}}
{}
\bibitem{Gerlich2012}
D. Gerlich, P. Jusko, Š. Roučka, I. Zymak, R. Plašil, and J. Glos\'{i}k, The
Astrophysical Journal \textbf{749}, 22 (2012). doi:\href
{https://doi.org/10.1088/0004-637X/749/1/22} {\nolinkurl
{10.1088/0004-637X/749/1/22}}
{}
\bibitem{Asvany2014}
O. Asvany, S. Brünken, L. Kluge, and S. Schlemmer, Appl. Phys. B
\textbf{114}, 203–211 (2014). doi:\href
{https://doi.org/10.1007/s00340-013-5684-y} {\nolinkurl
{10.1007/s00340-013-5684-y}}{}
\bibitem{Gunther2017}
A. Günther, P. Nieto, D. Müller, A. Sheldrick, D. Gerlich, and O. Dopfer,
Journal of Molecular Spectroscopy \textbf{332}, 8–15 (2017). doi:\href
{https://doi.org/10.1016/j.jms.2016.08.017} {\nolinkurl
{10.1016/j.jms.2016.08.017}}
{}
\bibitem{Kumar2018}
S. S. Kumar, F. Grussie, Y. V. Suleimanov, H. Guo, and H. Kreckel, Science
Advances \textbf{4}, eaar3417 (2018). doi:\href
{https://doi.org/10.1126/sciadv.aar3417} {\nolinkurl
{10.1126/sciadv.aar3417}}
{}
\bibitem{Jusko2019}
P. Jusko, S. Brünken, O. Asvany, S. Thorwirth, A. Stoffels, L. van der Meer,
G. Berden, B. Redlich, J. Oomens, and S. Schlemmer, Faraday Discuss.
\textbf{217}, 172–202 (2019). doi:\href
{https://doi.org/10.1039/C8FD00225H} {\nolinkurl {10.1039/C8FD00225H}}
{}
\bibitem{Campbell2020}
E. K. Campbell, Molecular Physics \textbf{118}, e1797918 (2020). doi:\href
{https://doi.org/10.1080/00268976.2020.1797918} {\nolinkurl
{10.1080/00268976.2020.1797918}}
{}
\bibitem{Luca2001}
A. Luca, S. Schlemmer, I. Čermák, and D. Gerlich, Review of Scientific
Instruments \textbf{72}, 2900–2908 (2001). doi:\href
{https://doi.org/10.1063/1.1373666} {\nolinkurl {10.1063/1.1373666}}
{}
\bibitem{Goebbert2009aip}
D. J. Goebbert, G. Meijer, and K. R. Asmis, AIP Conference Proceedings
\textbf{1104}, 22–29 (2009). doi:\href {https://doi.org/10.1063/1.3115605}
{\nolinkurl {10.1063/1.3115605}}{}
\bibitem{Wang2008}
X.-B. Wang and L.-S. Wang, Review of Scientific Instruments \textbf{79},
073108 (2008). doi:\href {https://doi.org/10.1063/1.2957610} {\nolinkurl
{10.1063/1.2957610}}
{}
\bibitem{Wolk2014}
A. B. Wolk, C. M. Leavitt, E. Garand, and M. A. Johnson, Accounts of
Chemical Research \textbf{47}, 202–210 (2014). doi:\href
{https://doi.org/10.1021/ar400125a} {\nolinkurl {10.1021/ar400125a}}
{}
\bibitem{Feraud2014}
G. Féraud, C. Dedonder, C. Jouvet, Y. Inokuchi, T. Haino, R. Sekiya, and
T. Ebata, The Journal of Physical Chemistry Letters \textbf{5}, 1236–1240
(2014). doi:\href {https://doi.org/10.1021/jz500478w} {\nolinkurl
{10.1021/jz500478w}}
{}
\bibitem{Jasik2013}
J. Jaš\'{i}k, J. Žabka, J. Roithová, and D. Gerlich, International Journal of Mass
Spectrometry \textbf{354-355}, 204–210 (2013). doi:\href
{https://doi.org/10.1016/j.ijms.2013.06.007} {\nolinkurl
{10.1016/j.ijms.2013.06.007}}
{}
\bibitem{Geistlinger2021}
K. Geistlinger, M. Fischer, S. Spieler, L. Remmers, F. Duensing, F. Dahlmann,
E. Endres, and R. Wester, Review of Scientific Instruments \textbf{92},
023204 (2021). doi:\href {https://doi.org/10.1063/5.0040866} {\nolinkurl
{10.1063/5.0040866}}
{}
\bibitem{Otto2009}
R. Otto, P. Hlavenka, S. Trippel, J. Mikosch, K. Singer, M. Weidemüller, and
R. Wester, Journal of Physics B: Atomic, Molecular and Optical Physics
\textbf{42}, 154007 (2009). doi:\href{https://doi.org/10.1088/0953-4075/42/15/154007} {\nolinkurl
{10.1088/0953-4075/42/15/154007}}
{}
\bibitem{Pedregosa2017}
J. Pedregosa-Gutierrez, C. Champenois, M. R. Kamsap, G. Hagel, M. Houssin,
and M. Knoop, Journal of Modern Optics \textbf{65}, 529–537 (2018).
doi:\href {https://doi.org/10.1080/09500340.2017.1408866} {\nolinkurl
{10.1080/09500340.2017.1408866}}
{}
\bibitem{Fanghanel2017}
S. Fanghänel, O. Asvany, and S. Schlemmer, Journal of Molecular
Spectroscopy \textbf{332}, 124–133 (2017). doi:\href
{https://doi.org/10.1016/j.jms.2016.12.003} {\nolinkurl
{10.1016/j.jms.2016.12.003}}
{}
\bibitem{Plasil2012}
R. Plašil, I. Zymak, P. Jusko, D. Mulin, D. Gerlich, and J. Glos\'{i}k, Phil. Trans.
R. Soc. A \textbf{370}, 5066–5073 (2012). doi:\href
{https://doi.org/10.1098/rsta.2012.0098} {\nolinkurl
{10.1098/rsta.2012.0098}}
{}
\bibitem{Asvany2008}
O. Asvany, O. Ricken, H. S. P. Müller, M. C. Wiedner, T. F. Giesen, and
S. Schlemmer, Phys. Rev. Lett. \textbf{100}, 233004 (2008). doi:\href
{https://doi.org/10.1103/PhysRevLett.100.233004} {\nolinkurl
{10.1103/PhysRevLett.100.233004}}
{}
\bibitem{Jusko2014}
P. Jusko, O. Asvany, A.-C. Wallerstein, S. Brünken, and S. Schlemmer, Phys.
Rev. Lett. \textbf{112}, 253005 (2014). doi:\href
{https://doi.org/10.1103/PhysRevLett.112.253005} {\nolinkurl
{10.1103/PhysRevLett.112.253005}}
{}\bibitem{Brunken2017}
S. Brünken, L. Kluge, A. Stoffels, J. Pérez-R\'{i}os, and S. Schlemmer, Journal of
Molecular Spectroscopy \textbf{332}, 67–78 (2017). doi:\href
{https://doi.org/10.1016/j.jms.2016.10.018} {\nolinkurl
{10.1016/j.jms.2016.10.018}}
{}
\bibitem{Ishiuchi2017}
S.-i. Ishiuchi, H. Wako, D. Kato, and M. Fujii, Journal of Molecular
Spectroscopy \textbf{332}, 45–51 (2017). doi:\href
{https://doi.org/10.1016/j.jms.2016.10.011} {\nolinkurl
{10.1016/j.jms.2016.10.011}}
{}
\bibitem{Choi2012}
C. M. Choi, D. H. Choi, N. J. Kim, and J. Heo, International Journal of Mass
Spectrometry \textbf{314}, 18–21 (2012). doi:\href
{https://doi.org/10.1016/j.ijms.2012.01.009} {\nolinkurl
{10.1016/j.ijms.2012.01.009}}
{}
\bibitem{Notzold2020}
M. Nötzold, S. Z. Hassan, J. Tauch, E. Endres, R. Wester, and
M. Weidemüller, Applied Sciences \textbf{10}, 5264 (2020). doi:\href
{https://doi.org/10.3390/app10155264} {\nolinkurl {10.3390/app10155264}}
{}
\bibitem{Lakhmanskaya2014}
O. Lakhmanskaya, T. Best, S. Kumar, E. Endres, D. Hauser, R. Otto,
S. Eisenbach, A. von Zastrow, and R. Wester, International Journal of Mass
Spectrometry \textbf{365-366}, 281–286 (2014). doi:\href
{https://doi.org/10.1016/j.ijms.2014.03.001} {\nolinkurl
{10.1016/j.ijms.2014.03.001}}
{}
\bibitem{Asvany2009}
O. Asvany and S. Schlemmer, International Journal of Mass Spectrometry
\textbf{279}, 147–155 (2009). doi:\href{https://doi.org/10.1016/j.ijms.2008.10.022} {\nolinkurl
{10.1016/j.ijms.2008.10.022}}
{}
\bibitem{Rajeevan2021}
G. Rajeevan, S. Mohandas, and S. S. Kumar, Physica Scripta \textbf{96},
124001 (2021). doi:\href {https://doi.org/10.1088/1402-4896/ac1472}
{\nolinkurl {10.1088/1402-4896/ac1472}}
{}
\bibitem{Kang2014}
H. Kang, G. Féraud, C. Dedonder-Lardeux, and C. Jouvet, The Journal of
Physical Chemistry Letters \textbf{5}, 2760–2764 (2014). doi:\href
{https://doi.org/10.1021/jz5012466} {\nolinkurl {10.1021/jz5012466}}
{}
\bibitem{Mark1996}
S. Mark and D. Gerlich, Chemical Physics \textbf{209}, 235–257 (1996).
doi:\href {https://doi.org/10.1016/0301-0104(96)00159-0} {\nolinkurl
{10.1016/0301-0104(96)00159-0}}
{}
\bibitem{Savic2020}
I. Savić, S. Schlemmer, and D. Gerlich, ChemPhysChem \textbf{21},
1429–1435 (2020). doi:\href {https://doi.org/10.1002/cphc.202000258}
{\nolinkurl {10.1002/cphc.202000258}}
{}
\bibitem{Haufler1996}
E. Haufler, “Niederenergetische Elektronen- und Protonen-Transfer Prozesse”,
PhD thesis (TU Chemnitz, 1996), \url
{https://nbn-resolving.org/urn:nbn:de:bsz:ch1-199700046}.
{}
\bibitem{Richthofen2002}
J. F. von Richthofen, “Untersuchung der Bildung, des Isotopenaustausches und
der Isomerisierung des Ionensystem HCO$^+$/HOC$^+$”, PhD thesis (TU
Chemnitz, 2002), \url
{https://nbn-resolving.org/urn:nbn:de:swb:ch1-200300302}.{}
\bibitem{Stahler2004}
S. W. Stahler and F. Palla, \emph{``The Formation of Stars''} (Wiley, 2004),
doi:\href {https://doi.org/10.1002/9783527618675} {\nolinkurl
{10.1002/9783527618675}}
{}
\bibitem{McGuire2022}
B. A. McGuire, The Astrophysical Journal Supplement Series \textbf{259}, 30
(2022). doi:\href {https://doi.org/10.3847/1538-4365/ac2a48} {\nolinkurl
{10.3847/1538-4365/ac2a48}}
{}
\bibitem{Herbst1973}
E. Herbst and W. Klemperer, The Astrophysical Journal \textbf{185},
505–534 (1973). doi:\href {https://doi.org/10.1086/152436} {\nolinkurl
{10.1086/152436}}
{}
\bibitem{Watson72b}
W. D. Watson and E. E. Salpeter, The Astrophysical Journal \textbf{175},
659 (1972). doi:\href {https://doi.org/10.1086/151587} {\nolinkurl
{10.1086/151587}}
{}
\bibitem{Wakelam15}
V. Wakelam, J.-C. Loison, E. Herbst, B. Pavone, A. Bergeat, K. Béroff,
M. Chabot, A. Faure, D. Galli, W. D. Geppert, D. Gerlich, P. Gratier,
N. Harada, K. M. Hickson, P. Honvault, S. J. Klippenstein, S. D. Le Picard,
G. Nyman, M. Ruaud, S. Schlemmer, I. R. Sims, D. Talbi, J. Tennyson, and
R. Wester, The Astrophysical Journal Supplement Series \textbf{217}, 20, 20
(2015). doi:\href {https://doi.org/10.1088/0067-0049/217/2/20} {\nolinkurl
{10.1088/0067-0049/217/2/20}}
{}
\bibitem{Sipila15b}
O. Sipilä, J. Harju, P. Caselli, and S. Schlemmer, Astron. Astrophys.
\textbf{581}, A122 (2015). doi:\href{https://doi.org/10.1051/0004-6361/201526468} {\nolinkurl
{10.1051/0004-6361/201526468}}
{}
\bibitem{Daly1960}
N. R. Daly, Review of Scientific Instruments \textbf{31}, 264–267 (1960).
doi:\href {https://doi.org/10.1063/1.1716953} {\nolinkurl
{10.1063/1.1716953}}
{}
\bibitem{wilcoxImprovedIonExtraction2002}
B. E. Wilcox, C. L. Hendrickson, and A. G. Marshall, Journal of the American
Society for Mass Spectrometry \textbf{13}, 1304–1312 (2002). doi:\href
{https://doi.org/10.1016/S1044-0305(02)00622-0} {\nolinkurl
{10.1016/S1044-0305(02)00622-0}}
{}
\bibitem{gibsonModellingMassAnalyzer2010}
J. R. Gibson, K. G. Evans, and S. Taylor, Journal of Mass Spectrometry
\textbf{45}, 364–371 (2010). doi:\href {https://doi.org/10.1002/jms.1720}
{\nolinkurl {10.1002/jms.1720}}
{}
\bibitem{SIMION}
\emph{SIMION}. \url {https://simion.com/}
{}
\bibitem{smigajSolvingBoundaryIntegral2015}
W. Śmigaj, T. Betcke, S. Arridge, J. Phillips, and M. Schweiger, ACM Trans.
Math. Softw. \textbf{41}, 6:1–6:40 (2015). doi:\href
{https://doi.org/10.1145/2590830} {\nolinkurl {10.1145/2590830}}
{}
\bibitem{betckeBemppclFastPython2021}
T. Betcke and M. W. Scroggs, Journal of Open Source Software \textbf{6},
2879 (2021). doi:\href {https://doi.org/10.21105/joss.02879} {\nolinkurl
{10.21105/joss.02879}}
{}
\bibitem{greengardFastAlgorithmParticle1987}L. Greengard and V. Rokhlin, Journal of Computational Physics \textbf{73},
325–348 (1987). doi:\href {https://doi.org/10.1016/0021-9991(87)90140-9}
{\nolinkurl {10.1016/0021-9991(87)90140-9}}
{}
\bibitem{SalomePlatform}
\emph{Salome Platform}. \url {https://www.salome-platform.org/}
{}
\bibitem{Gerlich1990}
D. Gerlich, The Journal of Chemical Physics \textbf{92}, 2377–2388 (1990).
doi:\href {https://doi.org/10.1063/1.457980} {\nolinkurl {10.1063/1.457980}}
{}
\bibitem{Honvault2012}
P. Honvault, M. Jorfi, T. González-Lezana, A. Faure, and L. Pagani, Phys.
Rev. Lett. \textbf{108}, 109903 (2012). doi:\href
{https://doi.org/10.1103/PhysRevLett.108.109903} {\nolinkurl
{10.1103/PhysRevLett.108.109903}}
{}
\bibitem{Grozdanov2012}
T. P. Grozdanov and R. McCarroll, The Journal of Physical Chemistry A
\textbf{116}, 4569–4577 (2012). doi:\href {https://doi.org/10.1021/jp210992g}
{\nolinkurl {10.1021/jp210992g}}
{}
\bibitem{Lezana2021}
T. González-Lezana, P. Hily-Blant, and A. Faure, The Journal of Chemical
Physics \textbf{154}, 054310 (2021). doi:\href
{https://doi.org/10.1063/5.0039629} {\nolinkurl {10.1063/5.0039629}}
{}
\bibitem{Beyer2019}
M. Beyer, N. Hölsch, J. Hussels, C.-F. Cheng, E. J. Salumbides,
K. S. E. Eikema, W. Ubachs, C. Jungen, and F. Merkt, Phys. Rev. Lett.
\textbf{123}, 163002 (2019). doi:\href
{https://doi.org/10.1103/PhysRevLett.123.163002} {\nolinkurl
{10.1103/PhysRevLett.123.163002}}{}
\bibitem{Gerlich2013}
D. Gerlich, R. Plašil, I. Zymak, M. Hejduk, P. Jusko, D. Mulin, and J. Glos\'{i}k,
The Journal of Physical Chemistry A \textbf{117}, 10068–10075 (2013).
doi:\href {https://doi.org/10.1021/jp400917v} {\nolinkurl
{10.1021/jp400917v}}
{}
\bibitem{Tellinghuisen2020}
J. Tellinghuisen, Analytical Chemistry \textbf{92}, 10863–10871 (2020).
doi:\href {https://doi.org/10.1021/acs.analchem.0c02178} {\nolinkurl
{10.1021/acs.analchem.0c02178}}
{}
\bibitem{Scott1997}
G. B. Scott, D. A. Fairley, C. G. Freeman, M. J. McEwan, P. Spanel, and
D. Smith, The Journal of Chemical Physics \textbf{106}, 3982–3987 (1997).
doi:\href {https://doi.org/10.1063/1.473116} {\nolinkurl {10.1063/1.473116}}
{}
\bibitem{Freeman1987}
C. G. Freeman, J. S. Knight, J. G. Love, and M. J. McEwan, International
Journal of Mass Spectrometry and Ion Processes \textbf{80}, 255–271 (1987).
doi:\href {https://doi.org/10.1016/0168-1176(87)87034-9} {\nolinkurl
{10.1016/0168-1176(87)87034-9}}
{}
\bibitem{Borodi2009}
G. Borodi, A. Luca, and D. Gerlich, International Journal of Mass
Spectrometry \textbf{280}, 218–225 (2009). doi:\href
{https://doi.org/10.1016/j.ijms.2008.09.004} {\nolinkurl
{10.1016/j.ijms.2008.09.004}}
{}
\bibitem{Schlemmer1999}
S. Schlemmer, T. Kuhn, E. Lescop, and D. Gerlich, International Journal of
Mass Spectrometry \textbf{185-187}, 589–602 (1999). doi:\href{https://doi.org/10.1016/S1387-3806(98)14141-6} {\nolinkurl
{10.1016/S1387-3806(98)14141-6}}
{}
\bibitem{Okumura1985}
M. Okumura, L. I. Yeh, and Y. T. Lee, The Journal of Chemical Physics
\textbf{83}, 3705–3706 (1985). doi:\href {https://doi.org/10.1063/1.449127}
{\nolinkurl {10.1063/1.449127}}
{}
\bibitem{Roithova2019}
J. Roithová, J. Jaš\'{i}k, J. J. Del Pozo Mellado, and D. Gerlich, Faraday Discuss.
\textbf{217}, 98–113 (2019). doi:\href {https://doi.org/10.1039/C8FD00196K}
{\nolinkurl {10.1039/C8FD00196K}}
{}
\bibitem{Benesch1980}
W. Benesch, D. Rivers, and J. Moore, J. Opt. Soc. Am. \textbf{70}, 792–799
(1980). doi:\href {https://doi.org/10.1364/JOSA.70.000792} {\nolinkurl
{10.1364/JOSA.70.000792}}
{}
\bibitem{Brummer2003}
M. Brümmer, C. Kaposta, G. Santambrogio, and K. R. Asmis, The Journal of
Chemical Physics \textbf{119}, 12700–12703 (2003). doi:\href
{https://doi.org/10.1063/1.1634254} {\nolinkurl {10.1063/1.1634254}}
{}
\bibitem{Chakrabarty2013}
S. Chakrabarty, M. Holz, E. K. Campbell, A. Banerjee, D. Gerlich, and
J. P. Maier, The Journal of Physical Chemistry Letters \textbf{4}, 4051–4054
(2013). doi:\href {https://doi.org/10.1021/jz402264n} {\nolinkurl
{10.1021/jz402264n}}
{}
\bibitem{Schmid2022}
P. C. Schmid, O. Asvany, T. Salomon, S. Thorwirth, and S. Schlemmer, The
Journal of Physical Chemistry A \textbf{126}, 8111–8117 (2022). doi:\href{https://doi.org/10.1021/acs.jpca.2c05767} {\nolinkurl
{10.1021/acs.jpca.2c05767}}
\end{thebibliography}
\end{document}